\documentclass[12pt]{conm-p-l}

\usepackage{amssymb}
\usepackage{epsfig}
\usepackage{cite}

\newtheorem{theorem}{Theorem}[section]

\newtheorem{proposition}[theorem]{Proposition}
\newtheorem{corollary}[theorem]{Corollary}

\theoremstyle{definition}

\newtheorem{example}[theorem]{Example}

\theoremstyle{remark}
\newtheorem{remark}[theorem]{Remark}

\numberwithin{equation}{section}

\newcommand{\cle}{\preceq}
\newcommand{\opl}{{\oplus}}

\newcommand{\rmin}{\mathbf{R}_{\min}}
\newcommand{\rmax}{\mathbf{R}_{\max}}
\newcommand{\suplim}{\sup\limits}
\newcommand{\sumlim}{\sum\limits}
\newcommand{\maxlim}{\max\limits}
\newcommand{\pd}[2]{\dfrac{\partial#1}{\partial#2}}
\newcommand{\CalB}{\mathcal{B}}

\newcommand{\maF}{\mathcal{F}}
\newcommand{\maD}{\mathcal{D}}

\newcommand{\cF}{{\mathcal F}}

\newcommand{\cN}{{\mathcal N}}

\newcommand{\0}{\mathbf{0}}
\newcommand{\1}{\mathbf{1}}
\newcommand{\cset}{\mathbf{C}}

\newcommand{\rset}{\mathbf{R}}
\newcommand{\maA}{\mathcal{A}}
\newcommand{\Log}{\mathop{\mathrm{Log}}}
\newcommand{\ovol}{\mathop{\mathrm{vol}}}

\def\C{\mathbf C}
\def\R{\mathbf R}
\def\maF{\mathcal F}
\def\maD{\mathcal D}

\def\cN{\mathcal N}
\def\Rmax{\rset_{\max}}

\begin{document}

\title{Tropical Mathematics, Idempotent Analysis, Classical Mechanics and Geometry}
\author{G.~L.~Litvinov}
\subjclass{Primary: 15A80, 46S19, 81Q20, 14M25, 16S80, 70H20, 14T05, 51P05, 52A20;
Secondary: 81S99, 52B70, 12K10, 46L65, 11K55, 28B10, 28A80, 28A25, 06F99, 16H99.}
\keywords{Tropical mathematics, idempotent
mathematics, idempotent functional analysis, classical mechanics, convex geometry, tropical
geometry, Newton polytopes.}

\thanks{This work is supported by the RFBR
grant 08--01--00601.}

\address{Independent University of Moscow,
Bol'shoi Vlasievskii per., 11, Moscow 119002, Russia}

\email{glitvinov@gmail.com}

\begin{abstract}
A very brief introduction to tropical and idempotent mathematics
(including idempotent functional analysis) is presented.
Applications to classical mechanics and geometry are especially
examined.
\end{abstract}

\maketitle

\hfill
\parbox{0.4\linewidth}{
{\bf To Mikhail Shubin with my admiration and gratitude}}

\tableofcontents

\normalsize

\section{Introduction}

Tropical mathematics can be treated  as  a result of a
dequantization of the traditional mathematics as  the Planck
constant  tends to zero  taking imaginary values. This kind of
dequantization is known as the Maslov dequantization and it leads
to a mathematics over tropical algebras like the max-plus algebra.
The so-called idempotent dequantization is a generalization of the
Maslov dequantization. The idempotent dequantization leads to
mathematics over idempotent semirings (exact definitions see below
in sections 2 and 3). For  example,
the field of real or complex numbers can be treated as a quantum
object whereas idempotent semirings can be examined  as
"classical" or "semiclassical" objects  (a semiring is called
idempotent  if the  semiring addition is idempotent, i.e. $x
\oplus x = x$), see~\cite{Li2005,LiMa95,LiMa96,LiMa98}.

Tropical algebras are idempotent semirings (and semifields). Thus tropical mathematics is a part of idempotent mathematics. Tropical
algebraic geometry can be treated as a result of the Maslov
dequantization applied to the traditional algebraic geometry (O.
Viro, G. Mikhalkin), see,
e.g.,~\cite{Iten2007,Mi2005,Mi2006,Vir2000,Vir2002,Vir2008}. There
are interesting relations and applications to the traditional
convex geometry.

    In the spirit of N.~Bohr's correspondence principle there
is a (heuristic)  correspondence  between important, useful, and
interesting  constructions  and  results over fields and similar
results  over  idempotent  semirings.  A systematic application of
this correspondence  principle  leads  to  a variety  of theoretical
and applied results~\cite{Li2005,LiMa95,LiMa96,LiMa98,LiMa2005}, see
Fig.~1.

\begin{figure}
\centering
\epsfig{file=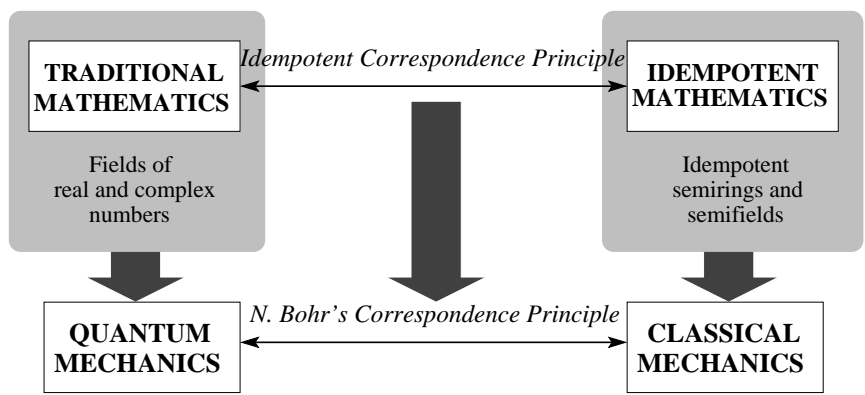,width=0.9\linewidth} \caption{Relations
between idempotent and traditional mathematics.}
\end{figure}

The history of the subject is discussed, e.g., in~\cite{Li2005}.
There is a large list of references.

\section{The Maslov dequantization}

Let $\R$ and $\C$ be the fields of real and complex numbers. The
so-called max-plus algebra $\R_{\max}= \R\cup\{-\infty\}$ is
defined by the operations $x\oplus y=\max\{x, y\}$ and $x\odot y=
x+y$.

The max-plus algebra can be treated as a result of the {\it Maslov
dequantization} of the semifield $\R_+$ of all nonnegative
numbers with the usual arithmetics. The change of variables
\begin{eqnarray*}
x\mapsto u=h\log x,
\end{eqnarray*}
where $h>0$, defines a map $\Phi_h\colon \R_+\to
\R\cup\{-\infty\}$, see Fig.~2. Let the addition and multiplication
operations be mapped from \markboth{G.L. Litvinov}{Tropical Mathematics, Idempotent Analysis, Classical Mechanics, and Geometry} $\R_+$ to $\R\cup\{-\infty\}$ by $\Phi_h$, i.e.\
let
\begin{eqnarray*}
u\oplus_h v = h \log({\mbox{exp}}(u/h)+{\mbox{exp}}(v/h)),\quad u\odot v= u+ v,\\
\mathbf{0}=-\infty = \Phi_h(0),\quad \mathbf{1}= 0 = \Phi_h(1).
\end{eqnarray*}

\begin{figure}
\noindent\epsfig{file=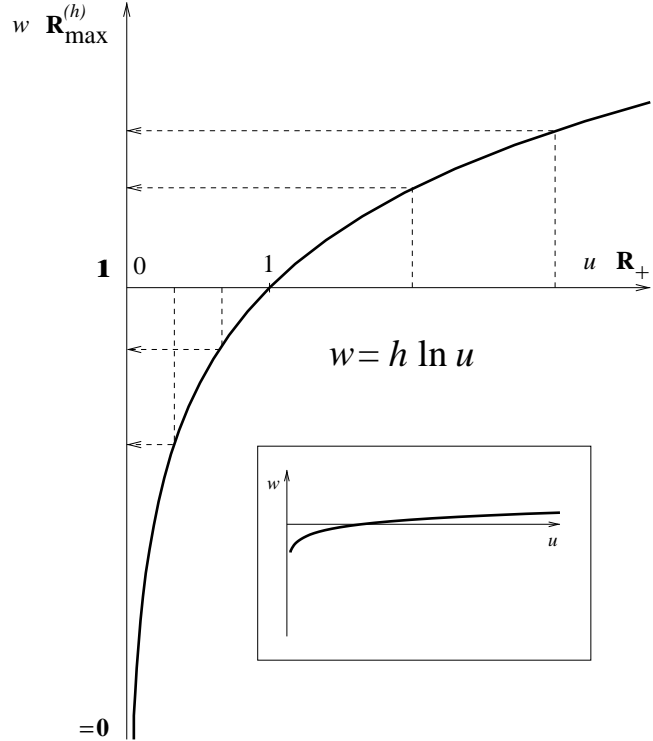,width=0.9\linewidth}
\vskip -3cm
\caption{Deformation of $\R_+$ to $\R^{(h)}$. Inset: the same for a
small value of $h$.}
\end{figure}

It can easily be checked that $u\oplus_h v\to \max\{u, v\}$ as
$h\to 0$. Thus we get the semifield $\R_{\max}$ (i.e.\ the
max-plus algebra) with zero $\mathbf{0}= -\infty$ and unit
$\mathbf{1}=0$ as a result of this deformation of the algebraic
structure~in~$\R_+$.

The semifield $\R_{\max}$ is a typical example of an {\it
 idempotent semiring}; this is a semiring with idempotent addition, i.e.,
 $x\oplus x = x$ for arbitrary element
 $x$ of this semiring.

 The semifield $\R_{\max}$ is also called a \emph{tropical
 algebra}. The semifield $\R^{(h)}=\Phi_h(\R_+)$ with operations
 $\oplus_h$ and $\odot$ (i.e.$+$) is called a \emph{subtropical
 algebra}.

 The semifield $\R_{\min}=\R\cup\{+\infty\}$
with operations $\oplus={\min}$ and $\odot=+$
$(\mathbf{0}=+\infty, \mathbf{1}=0)$ is isomorphic to $\R_{\max}$.

The analogy with quantization is obvious; the parameter $h$ plays
the role of the Planck constant. The map $x\mapsto|x|$ and the
Maslov dequantization for $\R_+$ give us a natural transition from
the field $\C$ (or $\R$) to the max-plus algebra $\R_{\max}$. {\it
We will also call this transition the Maslov dequantization}. In fact
the Maslov dequantization corresponds to the usual Schr\"odinger
dequantization but for  imaginary values of the Planck constant (see below).
The transition from numerical fields to the max-plus algebra
$\R_{\max}$ (or similar semifields) in mathematical constructions
and results generates the so called {\it tropical mathematics}.
The so-called {\it idempotent dequantization} is a generalization
of the Maslov dequantization; this is the transition from basic fields to idempotent semirings in mathematical constructions and results without any deformation. The idempotent dequantization generates
the so-called \emph{idempotent mathematics}, i.e. mathematics over
idempotent semifields and semirings.

{\bf Remark.} The term 'tropical' appeared in~\cite{Sim88} for a
discrete version of the max-plus algebra (as a suggestion of
Christian Choffrut). On the other hand V.~P.~Maslov used this term in
80s in his talks and works on economical applications of his
idempotent analysis (related to colonial politics). For the most
part of modern authors, 'tropical' means 'over $\R_{\max}$ (or
$\R_{\min}$)' and tropical algebras are $\R_{\max}$ and $\R_{\min}$.
The terms 'max-plus', 'max-algebra' and 'min-plus' are often used in
the same sense.
\medskip

\section{Semirings and semifields}

Consider a set $S$ equipped with two algebraic operations: {\it
addition} $\oplus$ and {\it multiplication} $\odot$. It is a {\it
semiring} if the following conditions are satisfied:
\begin{itemize}
\item the addition $\oplus$ and the multiplication $\odot$ are
associative; \item the addition $\oplus$ is commutative; \item the
multiplication $\odot$ is distributive with respect to the
addition $\oplus$:
\[x\odot(y\oplus z)=(x\odot y)\oplus(x\odot z)\]
and
\[(x\oplus y)\odot z=(x\odot z)\oplus(y\odot z)\]
for all $x,y,z\in S$.
\end{itemize}
A {\it unity} (we suppose that it exists) of a semiring $S$ is an element $\1\in S$ such that
$\1\odot x=x\odot\1=x$ for all $x\in S$. A {\it zero} (if it exists) of a
semiring $S$ is an element $\0\in S$ such that $\0\neq\1$ and
$\0\oplus x=x$, $\0\odot x=x\odot \0=\0$ for all $x\in S$. A
semiring $S$ is called an {\it idempotent semiring} if $x\oplus
x=x$ for all $x\in S$. A semiring $S$ with neutral element
$\1$ is called a {\it semifield} if every nonzero element of
$S$ is invertible with respect to the multiplication. The theory
of semirings and semifields is treated, e.g., in~\cite{Gol99}.
\medskip

\section{Idempotent analysis}

Idempotent analysis deals with functions taking their values in
an idempotent semiring and the corresponding function spaces.
Idempotent analysis was initially constructed by V.~P.~Maslov and
his collaborators and then developed by many authors. The subject
and applications are presented in the book of V.~N.~Kolokoltsov and
V.~P.~Maslov~\cite{KoMa97} (a version of this book in Russian was
published in 1994).

Let $S$ be an arbitrary semiring with idempotent addition $\oplus$
(which is always assumed to be commutative), multiplication
$\odot$, and unit $\1$. The set $S$ is supplied with
the {\it standard partial order\/}~$\cle$: by definition, $a \cle
b$ if and only if $a \oplus b = b$. If $S$ contains a zero element
$\0$, then all elements of $S$ are
nonnegative: $\0 \cle$ $a$ for all $a \in S$. Due to the existence
of this order, idempotent analysis is closely related to the
lattice theory, theory of vector lattices, and theory of ordered
spaces. Moreover, this partial order allows to model a number of
basic ``topological'' concepts and results of idempotent analysis
at the purely algebraic level; this line of reasoning was examined
systematically in~\cite{Li2005}--~\cite{LiSh2007}
and~\cite{CoGaQu2004}.

Calculus deals mainly with functions whose values are numbers. The
idempotent analog of a numerical function is a map $X \to S$,
where $X$ is an arbitrary set and $S$ is an idempotent semiring.
Functions with values in $S$ can be added, multiplied by each
other, and multiplied by elements of $S$ pointwise.

The idempotent analog of a linear functional space is a set of
$S$-valued functions that is closed under addition of functions
and multiplication of functions by elements of $S$, or an
$S$-semimodule. Consider, e.g., the $S$-semimodule $B(X, S)$ of
all functions $X \to S$ that are bounded in the sense of the
standard order on $S$.

If $S = \rmax$, then the idempotent analog of integration is
defined by the formula
$$
I(\varphi) = \int_X^{\oplus} \varphi (x)\, dx     = \sup_{x\in X}
\varphi (x),\eqno{(1)}
$$
where $\varphi \in B(X, S)$. Indeed, a Riemann sum of the form
$\sumlim_i \varphi(x_i) \cdot \sigma_i$ corresponds to the
expression $\bigoplus\limits_i \varphi(x_i) \odot \sigma_i =
\maxlim_i \{\varphi(x_i) + \sigma_i\}$, which tends to the
right-hand side of~(1) as $\sigma_i \to 0$. Of course, this is a
purely heuristic argument.

Formula~(1) defines the \emph{idempotent} (or \emph{Maslov})
\emph{integral} not only for functions taking values in $\rmax$,
but also in the general case when any of bounded (from above)
subsets of~$S$ has the least upper bound.

An \emph{idempotent} (or \emph{Maslov}) \emph{measure} on $X$ is
defined by the formula $m_{\psi}(Y) = \suplim_{x \in Y} \psi(x)$, where $\psi
\in B(X,S)$ is a fixed function. The integral with respect to this measure is defined
by the formula
$$
   I_{\psi}(\varphi)
    = \int^{\oplus}_X \varphi(x)\, dm_{\psi}
    = \int_X^{\oplus} \varphi(x) \odot \psi(x)\, dx
    = \sup_{x\in X} (\varphi (x) \odot \psi(x)).
    \eqno{(2)}
$$

Obviously, if $S = \rmin$, then the standard order is
opposite to the conventional order $\le$, so in this case
equation~(2) assumes the form
$$
   \int^{\oplus}_X \varphi(x)\, dm_{\psi}
    = \int_X^{\oplus} \varphi(x) \odot \psi(x)\, dx
    = \inf_{x\in X} (\varphi (x) \odot \psi(x)),
$$
where $\inf$ is understood in the sense of the conventional order
$\le$.
\medskip

\section{The superposition principle and linear problems}

Basic equations of quantum theory are linear; this is the
superposition principle in quantum mechanics. The Hamilton--Jacobi
equation, the basic equation of classical mechanics, is nonlinear
in the conventional sense. However, it is linear over the
semirings $\rmax$ and $\rmin$. Similarly, different versions of
the Bellman equation, the basic equation of optimization theory,
are linear over suitable idempotent semirings. This is
V.~P.~Maslov's idempotent superposition principle, see
\cite{Mas86,Mas87a,Mas87b}. For instance, the finite-dimensional
stationary Bellman equation can be written in the form $X = H
\odot X \oplus F$, where $X$, $H$, $F$ are matrices with
coefficients in an idempotent semiring $S$ and the unknown matrix
$X$ is determined by $H$ and $F$~\cite{Ca71, Ca79, BaCoOlQu92, Cu79, Cu95, GoMi79, GoMi2001}. In
particular, standard problems of dynamic programming and the
well-known shortest path problem correspond to the cases $S =
\rmax$ and $S =\rmin$, respectively. It is known that principal
optimization algorithms for finite graphs correspond to standard
methods for solving systems of linear equations of this type
(i.e., over semirings). Specifically, Bellman's shortest path
algorithm corresponds to a version of Jacobi's algorithm, Ford's
algorithm corresponds to the Gauss--Seidel iterative scheme,
etc.~\cite{Ca71, Ca79}.

The linearity of the Hamilton--Jacobi equation over $\rmin$ and
$\rmax$, which is the result of the Maslov dequantization of the
Schr{\"o}\-din\-ger equation, is closely related to the
(conventional) linearity of the Schr{\"o}\-din\-ger equation and
can be deduced from this linearity. Thus, it is possible to borrow
standard ideas and methods of linear analysis and apply them to a
new area.

Consider a classical dynamical system specified by the Hamiltonian
$$
   H = H(p,x) = \sum_{i=1}^N \frac{p^2_i}{2m_i} + V(x),
$$
where $x = (x_1, \dots, x_N)$ are generalized coordinates, $p =
(p_1, \dots, p_N)$ are generalized momenta, $m_i$ are generalized
masses, and $V(x)$ is the potential. In this case the Lagrangian
$L(x, \dot x, t)$ has the form
$$
   L(x, \dot x, t)
    = \sum^N_{i=1} m_i \frac{\dot x_i^2}2 - V(x),
$$
where $\dot x = (\dot x_1, \dots, \dot x_N)$, $\dot x_i = dx_i /
dt$. The value function $S(x,t)$ of the action functional has the
form
$$
   S = \int^t_{t_0} L(x(t), \dot x(t), t)\, dt,
$$
where the integration is performed along the factual trajectory of
the system.  The classical equations of motion are derived as the
stationarity conditions for the action functional (the Hamilton
principle, or the least action principle).

For fixed values of $t$ and $t_0$ and arbitrary trajectories
$x(t)$, the action functional $S=S(x(t))$ can be considered as a
function taking the set of curves (trajectories) to the set of
real numbers which can be treated as elements of  $\rmin$. In this
case the minimum of the action functional can be viewed as the
Maslov integral of this function over the set of trajectories or
an idempotent analog of the Euclidean version of the Feynman path
integral. The minimum of the action functional corresponds to the
maximum of $e^{-S}$, i.e. idempotent integral
$\int^{\oplus}_{\{paths\}} e^{-S(x(t))} D\{x(t)\}$ with respect to
the max-plus algebra $\rset_{\max}$. Thus the least action
principle can be considered as an idempotent version of the
well-known Feynman approach to quantum mechanics.  The
representation of a solution to the Schr{\"o}\-din\-ger equation
in terms of the Feynman integral corresponds to the
Lax--Ole\u{\i}nik solution formula for the Hamilton--Jacobi
equation.

Since $\partial S/\partial x_i = p_i$, $\partial S/\partial t =
-H(p,x)$, the following Hamilton--Jacobi equation holds:
$$
   \pd{S}{t} + H \left(\pd{S}{x_i}, x_i\right)= 0.\eqno{(3)}
$$

Quantization leads to the Schr\"odinger equation
$$
   -\frac{\hbar}i \pd{\psi}{t}= \widehat H \psi = H(\hat p_i, \hat x_i)\psi,
   \eqno{(4)}
$$
where $\psi = \psi(x,t)$ is the wave function, i.e., a
time-dependent element of the Hilbert space $L^2(\rset^N)$, and
$\widehat H$ is the energy operator obtained by substitution of
the momentum operators $\widehat p_i = {\hbar \over i}{\partial
\over \partial x_i}$ and the coordinate operators $\widehat x_i
\colon \psi \mapsto x_i\psi$ for the variables $p_i$ and $x_i$ in
the Hamiltonian function, respectively. This equation is linear in
the conventional sense (the quantum superposition principle). The
standard procedure of limit transition from the Schr\"odinger
equation to the Hamilton--Jacobi equation is to use the following
ansatz for the wave function:  $\psi(x,t) = a(x,t)
e^{iS(x,t)/\hbar}$, and to keep only the leading order as $\hbar
\to 0$ (the `semiclassical' limit).

Instead of doing this, we switch to imaginary values of the Planck
constant $\hbar$ by the substitution $h = i\hbar$, assuming $h >
0$. Thus the Schr\"odinger equation~(4) turns to an analog of the
heat equation:
$$
   h\pd{u}{t} = H\left(-h\frac{\partial}{\partial x_i}, \hat x_i\right) u,
   \eqno{(5)}
$$
where the real-valued function $u$ corresponds to the wave
function $\psi$. A similar idea (the switch to imaginary time) is
used in the Euclidean quantum field theory; let us remember that
time and energy are dual quantities.

Linearity of equation~(4) implies linearity of equation~(5). Thus
if $u_1$ and $u_2$ are solutions of~(5), then so is their linear
combination
$$
   u = \lambda_1 u_1 + \lambda_2 u_2.\eqno{(6)}
$$

Let $S = h \ln u$ or $u = e^{S/h}$ as in Section 2 above. It can
easily be checked that equation~(5) thus turns to
$$
   \pd{S}{t}= V(x) + \sum^N_{i=1} \frac1{2m_i}\left(\pd{S}{x_i}\right)^2
    + h\sum^n_{i=1}\frac1{2m_i}\frac{\partial^2 S}{\partial x^2_i}.
   \eqno{(7)}
$$
Thus we have a transition from (4) to (7) by means of the change of
variables $\psi = e^{S/h}$. Note that $|\psi| = e^{ReS/h}$ , where
Re$S$ is the real part  of $S$. Now let us consider $S$ as a real
variable. The equation (7) is nonlinear in the conventional sense.
However, if $S_1$ and $S_2$ are its solutions, then so is the
function
$$
   S = \lambda_1 \odot S_1 \opl_h \lambda_2\odot S_2
$$
obtained from~(6) by means of our substitution $S = h \ln u$. Here
the generalized multiplication $\odot$ coincides with the ordinary
addition and the generalized addition $\opl_h$ is the image of the
conventional addition under the above change of variables.  As $h
\to 0$, we obtain the operations of the idempotent semiring
$\rmax$, i.e., $\oplus = \max$ and $\odot = +$, and equation~(7)
turns to the Hamilton--Jacobi equation~(3), since the third term
in the right-hand side of equation~(7) vanishes.

Thus it is natural to consider the limit function $S = \lambda_1
\odot S_1 \oplus \lambda_2 \odot S_2$ as a solution of the
Hamilton--Jacobi equation and to expect that this equation can be
treated as linear over $\rmax$. This argument (clearly, a
heuristic one) can be extended to equations of a more general
form. For a rigorous treatment of (semiring) linearity for these
equations see, e.g., \cite{KoMa97,LiMa2005,Roub}. Notice that if
$h$ is changed to $-h$, then we have that the resulting
Hamilton--Jacobi equation is linear over $\rmin$.

The idempotent superposition principle indicates that there exist
important nonlinear (in the traditional sense) problems that are
linear over idempotent semirings. The idempotent linear functional
analysis (see below) is a natural tool for investigation of those
nonlinear infinite-dimensional problems that possess this
property.

\section{Convolution and the Fourier--Legendre transform}

Let $G$ be a group. Then the space $\CalB(G, \rset_{\max})$ of all
bounded functions $G\to\rset_{\max}$ (see above) is an idempotent
semiring with respect to the following analog $\circledast$ of the
usual convolution:
$$
   (\varphi(x)\circledast\psi)(g)=
    = \int_G^{\oplus} \varphi (x)\odot\psi(x^{-1}\cdot g)\, dx=
\sup_{x\in G}(\varphi(x)+\psi(x^{-1}\cdot g)).
$$
Of course, it is possible to consider other ``function spaces''
(and other basic semirings instead of $\rset_{\max}$).

Let $G=\rset^n$, where $\rset^n$ is considered as a topological
group with respect to the vector addition. The conventional
Fourier--Laplace transform is defined as
$$
   \varphi(x) \mapsto \tilde{\varphi}(\xi)
    = \int_G e^{i\xi \cdot x} \varphi (x)\, dx,\eqno{(8)}
$$
where $e^{i\xi \cdot x}$ is a character of the group $G$, i.e., a
solution of the following functional equation:
$$
   f(x + y) = f(x)f(y).
$$
The idempotent analog of this equation is
$$
   f(x + y) = f(x) \odot f(y) = f(x) + f(y),
$$
so ``continuous idempotent characters'' are linear functionals of
the form $x \mapsto \xi \cdot x = \xi_1 x_1 + \dots + \xi_n x_n$.
As a result, the transform in~(8) assumes the form
$$
   \varphi(x) \mapsto \tilde{\varphi}(\xi)
    = \int_G^\oplus \xi \cdot x \odot \varphi (x)\, dx
   = \sup_{x\in G} (\xi \cdot x + \varphi (x)).\eqno{(9)}
$$
The transform in~(9) is nothing but the {\it Legendre transform\/}
(up to some notation) \cite{Mas87b}; transforms of this kind
establish the correspondence between the Lagrangian and the
Hamiltonian formulations of classical mechanics. The Legendre
transform generates an idempotent version of harmonic analysis for
the space of convex functions, see, e.g., \cite{MaTi2003}.

Of course, this construction can be generalized to different
classes of groups and semirings. Transformations of this type
convert the generalized convolution $\circledast$ to the pointwise
(generalized) multiplication and possess analogs of some important
properties of the usual Fourier transform.

The examples discussed in this sections can be treated as
fragments of an idempotent version of the representation theory,
see, e.g., \cite{LiMaSh2002}. In particular, ``idempotent''
representations of groups can be examined as representations of
the corresponding convolution semirings (i.e. idempotent group
semirings) in semimodules.
\medskip

\section{Idempotent functional analysis}

Many other idempotent analogs may be given, in particular, for
basic constructions and theorems of functional analysis.
Idempotent functional analysis is an abstract version of
idempotent analysis. For the sake of simplicity take $S=\rmax$ and
let $X$ be an arbitrary set. The idempotent integration can be
defined by the formula (1), see above. The functional $I(\varphi)$
is linear over $S$ and its values correspond to limiting values of
the corresponding analogs of Lebesgue (or Riemann) sums. An
idempotent scalar product of functions $\varphi$ and $\psi$ is
defined by the formula
$$
\langle\varphi,\psi\rangle = \int^{\oplus}_X
\varphi(x)\odot\psi(x)\, dx = \sup_{x\in
X}(\varphi(x)\odot\psi(x)).
$$
So it is natural to construct idempotent analogs of integral
operators in the form
$$
\varphi(y) \mapsto (K\varphi)(x) = \int^{\oplus}_Y K(x,y)\odot
\varphi(y)\, dy = \sup_{y\in Y}\{K(x,y)+\varphi(y)\},\eqno(10)
$$
where $\varphi(y)$ is an element of a space of functions defined
on a set $Y$, and $K(x,y)$ is an $S$-valued function on $X\times
Y$. Of course, expressions of this type are standard in
optimization problems.\medskip

Recall that the definitions and constructions described above can
be extended to the case of idempotent semirings which are
conditionally complete in the sense of the standard order. Using
the Maslov integration, one can construct various function spaces
as well as idempotent versions of the theory of generalized
functions (distributions). For some concrete idempotent function
spaces it was proved that every `good' linear operator (in the
idempotent sense) can be presented in the form (10); this is an
idempotent version of the kernel theorem of L.~Schwartz; results
of this type were proved by V.~N.~Kolokoltsov, P.~S.~Dudnikov and
S.~N.~Samborski\u\i, I.~Singer, M.~A.~Shubin and others. So every
`good' linear functional can be presented in the form
$\varphi\mapsto\langle\varphi,\psi\rangle$, where
$\langle,\rangle$ is an idempotent scalar product.\medskip

In the framework of idempotent functional analysis results of this
type can be proved in a very general situation. In \cite{LiMaSh98,
LiMaSh99,LiMaSh2001,LiMaSh2002,LiSh2002,LiSh2007} an algebraic
version of the idempotent functional analysis is developed; this
means that basic (topological) notions and results are simulated
in purely algebraic terms (see below). The treatment covers the subject  from
basic concepts and results (e.g., idempotent analogs of the
well-known theorems of Hahn-Banach, Riesz, and Riesz-Fisher) to
idempotent analogs of A.~Grothendieck's concepts and results on
topological tensor products, nuclear spaces and operators.
Abstract idempotent versions of the kernel theorem are formulated. Note that
the transition from the usual theory to idempotent functional
analysis may be very nontrivial; for example, there are many
non-isomorphic idempotent Hilbert spaces. Important results on
idempotent functional analysis (duality and separation theorems)
were obtained by G.~Cohen, S.~Gaubert, and J.-P.~Quadrat.
Idempotent functional analysis has received much attention in the
last years, see,
e.g.,~\cite{AkGaKo2005, CoGaQu2004, GoMi79, GoMi2001, Gun98a, MaSa92, Shub92},~\cite{KoMa97}--~\cite{LiSh2007} and works
cited in~\cite{Li2005}. Elements of "tropical" functional analysis are presented in~\cite{KoMa97}. All the results presented in this section are proved in \cite{LiMaSh2001}  (subsections 7.1 -- 7.4) and in \cite{LiSh2007} (subsections 7.5 -- 7.10)

\subsection{Idempotent semimodules and idempotent linear spaces}

An additive semigroup $S$ with commutative addition $\oplus$ is
called an {\it idempotent semigroup} if the relation $x\oplus x=x$
is fulfilled for all elements $x\in S$. If $S$ contains a neutral
element, this element is denoted by the symbol $\0$. Any
idempotent semigroup is a partially ordered set with respect to
the following standard order: $x\preceq y$ if and only if $x\oplus
y=y$. It is obvious that this order is well defined and $x\oplus
y=\sup \{x, y\}$. Thus, any idempotent semigroup is an upper
semilattice; moreover, the concepts of idempotent semigroup and
upper semilattice coincide, see \cite{Bir}. An idempotent
semigroup $S$ is called $a$-{\it complete} (or {\it algebraically
complete}) if it is complete as an ordered set, i.e., if any
subset $X$ in $S$  has the least upper bound $\sup(X)$ denoted by
$\oplus X$ and the greatest lower bound $\inf (X)$ denoted by
$\wedge X$. This semigroup is called $b$-{\it complete} (or {\it
boundedly complete}), if any bounded above subset $X$ of this
semigroup (including the empty subset) has the least upper bound
$\oplus X$ (in this case, any nonempty subset $Y$ in $S$ has the
greatest lower bound $\wedge Y$ and $S$ in a lattice). Note that
any $a$-complete or $b$-complete idempotent semiring has the zero
element $\0$ that coincides with $\oplus\symbol{"1F}$, where
$\symbol{"1F}$ is the empty set. Certainly, $a$-completeness
implies the $b$-completeness. Completion by means of cuts
\cite{Bir} yields an embedding $S\to \widehat S$ of an arbitrary
idempotent semigroup $S$ into an $a$-complete idempotent semigroup
$\widehat S$ (which is called a {\it normal completion} of $S$);
in addition, $\widehat{\widehat S} = S$. The $b$-completion
procedure $S \to \widehat S_b$ is defined similarly: if $S \ni
\infty =\sup S$, then $\widehat S_b$ =$\widehat S$; otherwise,
$\widehat S =\widehat S_b \cup \{ \infty \}$. An arbitrary
$b$-complete idempotent semigroup  $S$ also may differ from
$\widehat S$ only by the element $\infty =\sup S$.

Let $S$ and $T$ be $b$-complete idempotent semigroups.
Then, a homomorphism $f : S\to T$ is said to be a $b$-{\it homomorphism} if
 $f (\oplus X) = \oplus f(X)$ for any bounded subset $X$ in $S$.
If the  $b$-homomorphism $f$ is extended to a homomorphism
$\widehat S\to \widehat T$ of the corresponding normal completions
and $f(\oplus X) = \oplus f(X)$ for all $X\subset S$, then $f$ is
said to be an $a$-{\it homomorphism}. An idempotent semigroup $S$
equipped with a topology such that the set $\{ s\in S\vert
s\preceq b\}$ is closed in this topology for any $b\in S$ is
called a {\it topological idempotent semigroup} $S$.

\begin {proposition} Let $S$ be an $a$-complete topological
idempotent semigroup and $T$ be a  $b$-complete topological
idempotent semigroup such that, for any nonempty subsemigroup $X$
in $T$, the element $\oplus X$ is contained in the topological
closure of $X$ in $T$. Then, a homomorphism $f : T\to S$ that maps
zero into zero is an $a$-homomorphism if and only if the mapping
$f$ is lower semicontinuous in the sense that the set $\{ t\in
T\vert f(t)\preceq s\}$ is closed in $T$ for any $s\in S$.
\end{proposition}
\medskip
An idempotent semiring $K$ is called $a$-{\it complete}
(respectively $b$-{\it complete}) if $K$ is an $a$-complete
(respectively $b$-complete) idempotent semigroup and, for any
subset (respectively, for any bounded subset) $X$ in $K$ and any
$k\in K$, the generalized distributive laws $k\odot(\oplus
X)=\oplus (k\odot X)$ and $(\oplus X)\odot k = \oplus(X\odot k)$
are fulfilled. Generalized distributivity implies that any
$a$-complete or $b$-complete idempotent semiring has a zero
element that coincides with $\oplus\symbol{"1F}$, where $
\symbol{"1F}$ is the empty set.

The set $\R(\max, +)$ of real numbers equipped with the idempotent
addition $\oplus=\max$ and multiplication $\odot=+$ is  an
idempotent semiring; in this case, $\1 = 0$. Adding the element
$\0=-\infty$ to this semiring, we obtain a $b$-complete semiring
$\R_{\max} = \R\cup \{-\infty\}$ with the same operations and the
zero element. Adding the element $+\infty$ to $\R_{\max}$ and
assuming that $\0\odot (+\infty)=\0$ and $x\odot (+\infty) =
+\infty$ for $x\neq\0$ and $x\oplus (+\infty) = +\infty$ for any
$x$, we obtain the $a$-complete idempotent semiring
$\widehat{\R}_{\max} = \R_{\max}\cup \{+\infty\}$. The standard
order on $\R(\max, +)$, $\R_{\max}$ and $\widehat{\R}_{\max}$
coincides with the ordinary order. The semirings $\R(\max,+)$ and
$\R_{\max}$ are semifields. On the contrary, an $a$-complete
semiring that does not coincide with $\{\0, \1\}$ cannot be a
semifield. An important class of examples is related to
(topological) vector lattices (see, for example, \cite{Bir} and
\cite{Scha}, Chapter 5). Defining the sum $x\oplus y$ as $\sup\{x,
y\}$ and the multiplication $\odot$ as the addition of vectors, we
can interpret the vector lattices as idempotent semifields. Adding
the zero element $\0$ to a complete vector lattice (in the sense
of \cite{Bir, Scha}), we obtain a $b$-complete semifield. If, in
addition, we add the infinite element, we obtain an $a$-complete
idempotent semiring (which, as an ordered set, coincides with the
normal completion of the original lattice).
\medskip

{\bf Important definitions.} Let $V$ be an idempotent semigroup and  $K$ be an
idempotent semiring. Suppose that a multiplication $k, x\mapsto k\odot x$
of all elements from $K$ by the elements from
$V$ is defined; moreover, this multiplication is associative
and distributive with respect
to the addition in $V$ and $\1\odot x = x$, $\0\odot x=\0$
for all $x\in V$. In this case, the semigroup $V$ is called an
{\it idempotent semimodule} (or simply, a {\it semimodule})
 over $K$. The element
 $\0_V\in V$ is called the {\it zero} of the semimodule $V$ if $k\odot\0_V=\0_V$
and $\0_V\oplus x = x$ for any $k\in K$ and $x\in V$.
Let $V$ be a semimodule over a $b$-complete idempotent semiring $K$.
This semimodule is called $b$-{\it complete} if it is $b$-complete as
an idempotent semiring and, for any bounded subsets $Q$ in $K$ and $X$ in  $V$,
the generalized distributive laws $(\oplus Q)\odot x = \oplus (Q\odot x)$ and
$k\odot (\oplus X) = \oplus (k\odot X)$ are fulfilled for all
$k\in K$ and $x\in X$. This semimodule is called $a$-{\it complete}
if it is $b$-complete and contains the element $\infty = \sup V$.

A semimodule $V$ over a $b$-complete semifield $K$ is said to be
an {\it idempotent} $a$-{\it space} ($b$-{\it space})
 if this semimodule is
$a$-complete (respectively, $b$-complete) and the equality
$(\wedge Q)\odot x = \wedge (Q\odot x)$ holds for any nonempty subset
$Q$ in $K$ and any $x\in V$, $x\neq \infty = \sup V$.
The normal completion $\widehat V$ of a $b$-space $V$ (as an idempotent semigroup) has the
structure of an idempotent $a$-space (and may differ from $V$ only
by the element $\infty= \sup V$).

Let $V$ and $W$ be idempotent semimodules over an idempotent semiring
$K$.  A mapping $p: V\to W$ is said to be {\it linear} (over $K$) if
$$
p(x\oplus y)=p(x)\oplus p(y) \mbox{ and } p(k\odot x)=k\odot p(x)
$$
for any $x, y\in V$ and $k\in K$. Let the semimodules $V$ and $W$ be
 $b$-complete. A linear mapping $p: V\to W$ is said to be $b$-{\it linear}
if it is a $b$-homomorphism of the idempotent semigroup; this mapping is said to be
$a$-{\it linear} if it can be extended to an $a$-homomorphism of the normal
completions $\widehat V$ and $\widehat W$. Proposition 7.1 (see above)
shows that $a$-linearity simulates (semi)continuity for
linear mappings. The normal completion $\widehat K$ of the
semifield $K$ is a semimodule over $K$. If $W= \widehat K$, then the
linear mapping $p$ is called a {\it linear functional}.

Linear, $a$-linear and $b$-linear mappings are also called {\it linear, a-linear} and {\it b-linear operators} respectively.

Examples of idempotent semimodules and spaces that are the most important
for analysis are either subsemimodules of topological vector lattices
\cite{Scha} (or coincide with them) or are dual to them, i.e., consist
of linear functionals subject to some regularity condition,
for example, consist of $a$-linear functionals. Concrete examples of
idempotent semimodules and spaces of functions (including spaces of
bounded, continuous, semicontinuous, convex, concave  and Lipschitz functions)
see in \cite{KoMa97, LiMaSh99, LiMaSh2001, LiSh2007} and below.
\medskip

\subsection{Basic results}

Let $V$ be an idempotent $b$-space over  a
$b$-complete semifield $K$,  $x\in \widehat V$. Denote by
$x^*$ the functional $V\to \widehat K$ defined by the formula
$x^* (y)=\wedge \{ k\in K \vert y\preceq k\odot x\}$,
where $y$ is an arbitrary fixed element from $V$.

\begin{theorem} For any $x\in \widehat V$ the functional $x^*$
is $a$-linear. Any nonzero $a$-linear functional $f$ on $V$ is
given by $f = x^*$ for a unique suitable element $x\in V$. If
$K\neq \{\0, \1\}$, then $x=\oplus\{ y\in V\vert f(y)\preceq\1\}$.
\end{theorem}

Note that results of this type obtained earlier concerning the structure of
linear functionals cannot be carried over to subspaces and
subsemimodules.

A subsemigroup $W$ in $V$ closed with respect to the
multiplication by an arbitrary element from $K$ is called a
$b$-{\it subspace} in $V$ if the imbedding $W\to V$ can be extended to a
$b$-linear mapping. The following result is obtained from Theorem 7.2
and is the idempotent version of the Hahn--Banach theorem.

\begin{theorem} Any $a$-linear functional defined on a
$b$-subspace $W$ in $V$ can be extended to an $a$-linear
functional on $V$. If $x, y\in V$ and $x\neq y$, then there exists
an $a$-linear functional $f$ on $V$ that separates the elements
$x$ and $y$, i.e., $f(x)\neq f(y)$.
\end{theorem}

The following statements are easily derived from the definitions and
can be regarded as the analogs of the well-known results of the traditional functional analysis (the Banach--Steinhaus and the closed-graph theorems).

\begin{proposition}  Suppose that $P$ is a family of $a$-linear
mappings of an $a$-space $V$ into an $a$-space $W$ and the mapping
$p : V\to W$ is the pointwise sum of the mappings of this family,
i.e., $p(x) =\sup \{ p_{\alpha} (x)\vert p_{\alpha} \in P\}$.
Then the mapping $p$ is $a$-linear.
\end{proposition}

\begin{proposition} Let $V$ and $W$ be  $a$-spaces. A linear
mapping $p : V \to W$ is $a$-linear if and only if its graph
$\Gamma$ in $V\times W$ is closed with respect to passing
to sums
(i.e., to  least upper bounds)
of its arbitrary subsets.
\end{proposition}

In \cite{CoGaQu2004} the basic results were generalized for the
case of semimodules over the so-called reflexive $b$-complete semirings.
\medskip

\subsection{Idempotent $b$-semialgebras}

Let $K$ be a $b$-complete semifield and $A$ be an
idempotent $b$-space over $K$ equipped with the structure of a semiring
compatible with the multiplication $K\times A\to A$ so that
the associativity of the multiplication is preserved. In this case,
$A$ is called an {\it idempotent $b$-semialgebra} over~$K$.

\begin{proposition} For any invertible element $x\in A$ from the
$b$-semialgebra $A$ and any element $y\in A$, the equality $x^*
(y) = \1^* (y\odot x^{-1})$ holds, where $\1\in A$.
\end{proposition}

The mapping $A\times A\to \widehat K$ defined by the formula
$(x, y) \mapsto \langle x, y\rangle = \1^* (x\odot y)$
is called the  {\it canonical scalar product} (or simply {\it scalar product}).
The basic properties of the scalar product are easily derived from
Proposition 7.6 (in particular, the scalar product is commutative
if the $b$-semialgebra $A$ is commutative). The following theorem is an idempotent version of the Riesz--Fisher theorem.

\begin{theorem} Let a $b$-semialgebra $A$ be a semifield.
Then any nonzero $a$-linear functional $f$ on $A$ can be represented
as $f(y) = \langle y, x\rangle$, where $x\in A$, $x\neq \0$ and
$\langle \cdot, \cdot\rangle$ is the canonical scalar product on
$A$.
\end{theorem}

\begin{remark} Using the completion precedures, one can extend all
the results obtained to the case of incomplete semirings, spaces,
 and semimodules, see~\cite{LiMaSh2001}.
\end{remark}

\begin{example} Let $ B (X)$ be a set of all bounded
functions with values belonging to $\R(\max, +)$ on an arbitrary
set $X$ and let $\widehat B (X) =  B (X)\cup \{\0\}$. The
pointwise idempotent addition of functions $(\varphi_1 \oplus
\varphi_2) (x) = \varphi_1(x)\oplus \varphi_2(x)$ and the
multiplication $(\varphi_1\odot \varphi_2)(x) =
(\varphi_1(x))\odot ( \varphi_2(x))$ define on $\widehat B (X)$
the structure of a $b$-semialgebra over the $b$-complete semifield
$\R_{\max}$. In this case, $\1^* (\varphi) =\sup_{x\in X} \varphi
(x)$ and the scalar product is expressed in terms of idempotent
integration: $\langle\varphi_1, \varphi_2\rangle  = \sup_{x\in X}
(\varphi_1 (x)\odot\varphi_2 (x)) = \sup_{x\in X} (\varphi_1 (x) +
\varphi_2 (x)) =\int\limits^{\oplus}_X(\varphi_1 (x)\odot\varphi_2
(x))\; dx$. Scalar products of this type were systematically used
in idempotent analysis. Using Theorems 7.2 and 7.7, one can easily
describe $a$-linear functionals on idempotent spaces in terms of
idempotent measures and integrals.
\end{example}

\begin{example} Let $X$ be a linear space in the traditional sense.
The idempotent semiring (and linear space over $\R(\max, +)$) of convex functions
Conv$(X, \R)$
is $b$-complete but it is not a $b$-semialgebra
over the
semifield
$K = \R(\max, +)$.
 Any nonzero $a$-linear functional~$f$
on
Conv$(X, \R)$
has the form
$$
{\varphi}\mapsto f({\varphi})
=\sup_x\{{\varphi}(x)+\psi(x)\}
=\int^\oplus_X{\varphi(x)}\odot\psi(x)\,dx,
$$
where~$\psi$ is a concave function, i.e., an element of the
idempotent space Conc$(X, \R)$ = - Conv$(X, \R)$.
\end{example}
\medskip

\subsection{Linear operator, $b$-semimodules and subsemimodules}

In what follows, we suppose that all semigroups, semirings, semifields, semimodules, and spaces are idempotent unless
otherwise specified. We fix a basic semiring $K$ and examine semimodules and subsemimodules over $K$. We suppose that every
linear functional takes it values in the basic semiring.

Let $V$ and $W$ be $b$-complete semimodules over a $b$-complete
semiring $K$. Denote by $L_b(V,W)$ the set of all $b$-linear
mappings from $V$ to $W$. It is easy to check that $L_b(V,W)$
is an idempotent semigroup with respect to the pointwise
addition of operators; the composition (product) of $b$-linear
operators is also a $b$-linear operator, and therefore
the set $L_b(V,V)$
is an idempotent semiring with respect to these operations,
see, e.g., \cite{LiMaSh2001}. The following proposition can be treated
as a version of the Banach--Steinhaus theorem in idempotent
analysis (as well as Proposition 7.4 above).

\begin{proposition} Assume that $S$ is a subset in $L_b(V,W)$ and the set
$\{g(v)\mid g\in S\}$ is bounded in $W$ for every element
$v\in V$; thus the element $f(v)$ = $\sup_{g\in S}{g(v)}$
exists, because the semimodule $W$ is $b$-complete. Then the
mapping $v\mapsto f(v)$ is a $b$-linear operator, i.e., an
element of $L_b(V,W)$. The subset $S$ is bounded; moreover, $\sup S = f$.
\end{proposition}

\begin{corollary} The set $L_b(V,W)$ is a $b$-complete
idempotent semigroup
with respect to the (idempotent) pointwise addition of
operators. If $V = W$, then $L_b(V,V)$ is a $b$-complete idempotent
semiring with respect to the operations of pointwise addition and
composition of operators.
\end{corollary}

\begin{corollary} A subset $S$ is bounded in
$L_b(V,W)$  if and only if the set $\{g(v)\mid g\in S\}$ is
bounded in the semimodule $W$ for every element $v\in V$.
\end{corollary}

A subset of an idempotent semimodule is called a {\it subsemimodule}
if it is closed under addition and multiplication by scalar coefficients.
A subsemimodule $V$ of a $b$-complete semimodule $W$ is {\it b-closed}
if $V$ is closed under sums of any subsets of $V$
that are bounded in $W$.
A subsemimodule of a $b$-complete semimodule is called a
{\it b-subsemimodule} if the corresponding embedding is a $b$-homomorphism.
It is easy to see that each $b$-closed subsemimodule is a $b$-subsemimodule,
but the converse is not true.
The main feature of $b$-subsemimodules is that restrictions of $b$-linear
operators and functionals to these semimodules are $b$-linear.
\medskip

{\it The following definitions are very important} for our
purposes. Assume that $W$ is an idempotent $b$-complete semimodule over a
$b$-complete idempotent semiring $K$ and $V$ is a subset of $W$ such
that $V$ is closed under multiplication by scalar coefficients
and is an upper semilattice with respect to the order induced
from $W$. Let us define an addition operation in $V$ by the
formula $x\oplus y = \sup\{ x, y \}$, where $\sup$ means the least
upper bound in $V$. If $K$ is a semifield, then $V$ is a
semimodule over $K$ with respect to this addition.

For an arbitrary $b$-complete semiring $K$, we will say that
$V$ is a {\it quasisubsemimodule} of $W$ if $V$ is a
semimodule with respect to this addition (this means that the
corresponding distribution laws hold).

Recall that the simbol $\wedge$ means the greatest lower bound (see Subsection 7.1 above). A quasisubsemimodule $V$ of an idempotent $b$-complete semimodule
$W$ is called a $\wedge$-{\it subsemimodule} if it contains
$\0$ and is closed under the operations of taking infima (greatest lower
bounds) in $W$. It is easy to check  that {\it each $\wedge$-subsemimodule is a $b$-complete
semimodule}.

Note that quasisubsemimodules and $\wedge$-subsemimodules
may fail to be subsemimodules, because only the order is induced
and not the
corresponding addition (see Example 7.18 below).

Recall that idempotent semimodules over semifields
are {\it idempotent spaces}. In idempotent mathematics, such spaces
are analogs of traditional linear (vector) spaces over fields.
In a similar way we use the corresponding terms like
{\it b-spaces, b-subspaces, b-closed subspaces},
$\wedge$-{\it subspaces}, etc.

Some examples are presented below.
\medskip

\subsection{Functional semimodules}

Let $X$ be an arbitrary nonempty set and $K$ be an idempotent semiring.
By $K(X)$ denote the semimodule of all mappings (functions) $X \to K$
endowed with the pointwise operations. By $K_b(X)$ denote the subsemimodule
of $K(X)$ consisting of  all bounded
mappings. If $K$ is a $b$-complete semiring, then $K(X)$ and $K_b(X)$ are
$b$-complete semimodules. Note that $K_b(X)$ is a $b$-subsemimodule but not
a $b$-closed subsemimodule of $K(X)$. Given a point $x\in X$, by
$\delta _x$ denote the functional on $K(X)$ that maps
$f$ to $f(x)$. It can easily be checked that the functional
$\delta _x$ is $b$-linear on~$K(X)$.

We say that a quasisubsemimodule of $K(X)$ is an (idempotent)
{\it functional semimodule} on the set $X$. An idempotent functional
semimodule in $K(X)$ is called {\it b-complete} if it is a $b$-complete
semimodule.

A functional semimodule $V\subset K(X)$ is called a {\it functional
b-semimodule} if it is a b-subsemimodule of $K(X)$; a functional
semimodule $V\subset K(X)$ is called a {\it functional $\wedge$-semimodule}
if it is a $\wedge$-subsemimodule of $K(X)$.

In general, a functional of the form $\delta _x$ on a functional semimodule
is not even linear, much less $b$-linear
(see Example~7.18 below). However, the following
proposition holds, which is a direct consequence of our definitions.

\begin{proposition} An arbitrary $b$-complete functional semimodule
$W$ on a set $X$ is a $b$-subsemimodule of $K(X)$ if and only if
each functional of the form $\delta _x$ (where $x\in X$) is
$b$-linear on $W$.
\end{proposition}

\begin{example} The semimodule $K_b(X)$ (consisting of all bounded
mappings from an arbitrary set $X$ to a
$b$-complete idempotent semiring $K$)
is a functional $\wedge$-semimodule. Hence it is a $b$-complete
semimodule over $K$. Moreover, $K_b(X)$ is a $b$-subsemimodule of the
semimodule $K(X)$ consisting of all mappings $X\to K$.
\end{example}

\begin{example} If $X$ is a finite set consisting of $n$
elements ($n>0$), then $K_b(X) = K(X)$ is an
``$n$-dimensional'' semimodule over $K$; it is denoted by
$K^n$. In particular, $\R_{max}^n$ is an idempotent space over
the semifield $\R_{max}$, and $\widehat{\R}_{\max}^n$
is a semimodule over the semiring $\widehat{\R}_{\max}$. Note
that $\widehat{\R}_{\max}^n$ can be treated as a space
over the semifield $\R_{max}$. For example, the semiring
 $\widehat{\R}_{\max}$ can be treated as a space
(semimodule) over~$\Rmax$.
\end{example}

\begin{example} Let $X$ be a topological space. Denote by
$USC(X)$ the set of all upper semicontinuous functions with
values in $\R_{\max}$. By definition, a function $f(x)$ is upper
semicontinuous if the set $X_s =\{ x\in X\mid f(x)\geq s\}$
is closed in $X$ for every element $s\in \R_{\max}$ (see, e.g.,
\cite{LiMaSh2001}, Sec.~2.8). If a family $\{f_{\alpha}\}$ consists of
upper semicontinuous (e.g., continuous) functions and
$f(x) = \inf_{\alpha} f_{\alpha} (x)$, then $f(x) \in USC(X)$.
It is easy to check that $USC(X)$  has a natural structure
of an idempotent space over $\Rmax$. Moreover, $USC(X)$ is a functional
$\wedge$-space on $X$ and a b-space. The subspace
$USC(X)\cap K_b(X)$ of $USC(X)$ consisting of bounded (from
above)
functions has the same properties.
\end{example}

\begin{example}  Note that an idempotent functional semimodule
(and even a functional $\wedge$-semimodule) on a set $X$ is not
necessarily a subsemimodule of $K(X)$. The simplest example is the
functional space (over $K=\Rmax$) Conc(${\R}$) consisting of all
concave functions on $\R$ with values in $\Rmax$. Recall that a
function $f$ belongs to Conc(${\R}$) if and only if the subgraph
of this function is convex, i.e., the formula $f(ax+(1-a)y)\ge
af(x)+ (1-a)f(y)$ is valid for $0\le a\le 1$. The basic operations
with $\0\in \Rmax$ can be defined in an obvious way. If $f,g
\in$Conc$({\R})$, then denote by $f\oplus g$ the sum of these
functions in Conc$({\R})$. The subgraph of $f\oplus g$ is the
convex hull of the subgraphs of $f$ and $g$. Thus $f\oplus g$ does
not coincide with the pointwise sum (i.e., $\max\{f(x), g(x)\}$).
\end{example}

\begin{example} Let $X$ be a nonempty metric space with a
fixed metric $r$. Denote by Lip$(X)$ the set of all functions
defined on $X$ with values in $\Rmax$ satisfying the
following {\it Lipschitz condition}:
$$
\mid f(x)\odot (f(y))^{-1}\mid   =  \mid f(x) - f(y)\mid  \leq  r(x, y),
$$
where $x$, $y$ are arbitrary elements of $X$.  The set Lip$(X)$
consists of continuous real-valued functions (but not all of
them!) and (by definition) the function equal to $-\infty = \0$ at
every point $x\in X$. The set Lip$(X)$ has the structure of an
idempotent space over the semifield $\Rmax$. Spaces of the form
Lip$(X)$ are said to be  {\it Lipschitz spaces}. These spaces are
$b$-subsemimodules in $K(X)$.
\end{example}
\medskip

\subsection{Integral representations of
linear operators in functional semimodules}

Let $W$ be an idempotent $b$-complete semimodule over a $b$-complete
semiring $K$ and $V\subset K(X)$ be a $b$-complete functional
semimodule on $X$. A mapping $A:V\to W$ is called an {\it integral operator}
or an operator with an {\it integral representation} if there exists
a mapping $k:X\to W$, called the {\it integral kernel} (or {\it kernel})
{\it of the operator} $A$, such that
$$
Af = \sup_{x\in X}{(f(x)\odot k(x))}.
\eqno{(11)}
$$
In idempotent analysis, the right-hand side of formula (11) is often
written as $\int_X^{\oplus}f(x)\odot k(x) dx$.
Regarding the kernel $k$, it is assumed that the set
$\{f(x)\odot k(x)|x\in X\}$ is bounded in $W$ for all $f\in V$ and $x\in X$.
We denote the set of all functions with this property by
${\text{\rm kern}}_{V,W}(X)$. In particular, if $W=K$ and $A$ is a functional,
then this functional is called {\it integral}. Thus each integral
functional can be presented in the form of a ``scalar product''
$f \mapsto \int_X^{\oplus} f(x) \odot k(x)\; dx$, where
$k(x)\in K(X)$; in idempotent analysis, this situation is
standard.

Note that a functional of the form $\delta_y$ (where $y\in X$)
is a typical integral functional; in this case,
$k(x) = \bold1$ if $x = y$ and $k(x) = \0$ otherwise.

We call a functional semimodule $V\subset K(X)$ {\it
nondegenerate} if for every point $x\in X$ there exists a function
$g\in V$ such that $g(x)=\bold1$, and {\it admissible} if for
every function $f\in V$ and every point $x\in X$ such that
$f(x)\neq \0$ there exists a function $g\in V$ such that
$g(x)=\bold1$ and~$f(x)\odot g <f$.

Note that all idempotent functional semimodules over semifields
are admissible (it is sufficient to set $g = f(x)^{-1}\odot f$).

\begin{proposition} Denote by $X_V$  the subset of $X$ defined by the
formula $X_V=\{ x\in X\mid\; \exists f\in V: f(x)=\bold1 \}$. If the semimodule
$V$ is admissible, then the restriction to $X_V$ defines an
embedding $i:V\to K(X_V)$ and its image $i(V)$ is admissible and
nondegenerate.

If a mapping $k:X\to W$ is a kernel of a mapping $A:V\to W$, then the
mapping $k_V:X\to W$ that is equal to $k$ on $X_V$ and equal to $\0$
on $X\smallsetminus~X_V$ is also a kernel of~$A$.

A mapping $A:V\to W$ is integral if and only if the mapping
 $i_{-1}A:i(A)\to W$ is integral.
\end{proposition}

In what follows, $K$ always denotes a fixed $b$-complete idempotent
(basic) semiring.  If an operator has an integral representation,
this representation may not be unique. However, if the semimodule
$V$ is nondegenerate, then the set of all kernels of a fixed
integral operator is bounded with respect to the natural order
in the set of all kernels and is closed under the supremum
operation applied to its arbitrary subsets. In particular,
{\it any integral
operator defined on a nondegenerate functional semimodule has
a unique maximal kernel}.

An important point is that an integral operator is not necessarily $b$-linear
and even linear except when $V$ is a $b$-subsemimodule of $K(X)$
(see Proposition~7.21 below).

If $W$ is a functional semimodule on a nonempty set $Y$,
then an integral kernel $k$ of an operator $A$ can be naturally identified
with the function on $X\times Y$ defined by the formula
$k(x,y)=(k(x))(y)$. This function will also be called an
{\it integral kernel} (or {\it kernel}) of the operator $A$.
As a result, the set  ${\text{\rm kern}}_{V,W}(X)$  is identified with the set
${\text{\rm kern}}_{V,W}(X,Y)$ of all mappings $k: X\times Y\to K$ such that for every
point $x\in X$ the mapping $k_x: y\mapsto k(x,y)$ lies in $W$ and for
every $v\in V$ the set $\{v(x)\odot k_x|x\in X\}$ is bounded in $W$.
Accordingly, the set of all integral kernels of $b$-linear
operators can be embedded into~${\text{\rm kern}}_{V,W}(X,Y)$.

If $V$ and $W$ are functional $b$-semimodules on $X$ and $Y$,
respectively, then the set of all kernels of $b$-linear operators can
be identified with ${\text{\rm kern}}_{V,W}(X,Y)$
and the following formula holds:
$$
Af(y)=\sup_{x\in X}{(f(x)\odot k(x,y))}=\int_X^{\oplus} f(x)
\odot k(x,y) dx.
\eqno{(12)}
$$
This formula coincides with the usual definition of an
integral representation of an operator. Note that formula (11) can
be rewritten in the form
$$
Af=\sup_{x\in X}{(\delta _x(f)\odot k(x))}. \eqno{(13)}
$$

\begin{proposition} An arbitrary b-complete functional semimodule
$V$ on a nonempty set $X$ is a functional b-se\-mi\-mo\-du\-le on
$X$ (i.e., a b-sub\-semi\-mo\-du\-le of $K(X)$) if and only if
all integral operators defined on $V$ are b-linear.
\end{proposition}

The following notion (definition) is especially important
for our purposes. Let $V\subset K(X)$ be a $b$-complete functional
semimodule over a $b$-complete idempotent semiring $K$. We
say that the {\it kernel theorem} holds for the
semimodule $V$ if every $b$-linear mapping from $V$ into an
arbitrary $b$-complete semimodule over $K$ has an integral
representation.

\begin{theorem} Assume that  a b-complete
semimodule $W$ over a b-complete semiring $K$ and an
admissible functional $\wedge$-semimodule $V\subset K(X)$ are given.
Then every b-linear operator $A:V\to W$ has an integral
representation of the form $(11)$. In particular, if $W$ is
a functional b-semimodule on a set $Y$, then the operator
$A$ has an integral representation of the form $(12)$. Thus
for the semimodule $V$ the kernel theorem holds.
\end{theorem}

\begin{remark} Examples of admissible functional
$\wedge$-semimodules (and $\wedge$-spaces) appearing in
Theorem~7.22 are presented above, see, e.g., examples 7.15 --7.17.
Thus for these functional semimodules and spaces $V$ over $K$, the
kernel theorem holds and every $b$-linear  mapping $V$ into an
arbitrary $b$-complete semimodule $W$ over $K$ has an integral
representation (12). Recall that every functional space over a
$b$-complete semifield is admissible, see above.
\end{remark}
\medskip

\subsection{Nuclear operators and their integral representations}

Let us introduce some important definitions. Assume that
$V$ and $W$ are $b$-complete semimodules. A mapping $g: V\to
W$ is called {\it one-dimensional} (or a {\it mapping of
rank} 1) if it is of the form $v\mapsto \phi (v)\odot w$,
where $\phi$ is a $b$-linear functional on $V$ and $w\in W$.
A mapping $g$ is called {\it b-nuclear} if it is the sum (i.e., supremum)
of a bounded set of one-dimensional mappings. Since every
one-dimensional mapping is $b$-linear (because the functional
$\phi$ is $b$-linear), {\it every b-nuclear operator is b-linear}
(see Corollary~7.12 above). Of course, $b$-nuclear mappings are closely
related to tensor products of idempotent semimodules, see~\cite{LiMaSh99}.

By $\phi\odot w$ we denote the one-dimensional operator
$v\mapsto\phi(v)\odot w$. In fact, this is an element of
the corresponding tensor product.

\begin{proposition} The composition (product) of a
b-nuclear and a
b-linear mapping or of a b-linear and a b-nuclear mapping
is a b-nuclear operator.
\end{proposition}

\begin{theorem} Assume that $W$ is a b-complete semimodule over a
b-complete semiring $K$ and $V\subset K(X)$ is a functional b-semimodule.
If every b-linear functional on $V$ is integral, then a b-linear
operator $A:V\to W$ has an integral representation if and only if
it is b-nuclear.
\end{theorem}
\medskip

\subsection{The $b$-approximation property and $b$-nuclear
 semimodules and spaces}

We say that a $b$-complete semimodule $V$ has the {\it
b-approximation property} if the identity operator id:$V\to V$ is
$b$-nuclear (for a treatment of the approximation property for
locally convex spaces in the traditional functional
analysis, see \cite{Scha}).

Let $V$ be an arbitrary $b$-complete semimodule over a $b$-complete
idempotent semiring $K$. We call this semimodule a {\it b-nuclear
semimodule} if any $b$-linear mapping of $V$ to an arbitrary
$b$-complete semimodule $W$ over $K$ is a $b$-nuclear operator. Recall
that, in the traditional functional analysis, a locally convex space
is nuclear if and only if all continuous linear mappings of this
space to any Banach space are nuclear operators, see~\cite{Scha}.

\begin{proposition} Let $V$ be an arbitrary
b-complete semimodule over a b-complete semiring $K$. The
following statements are equivalent:
\smallskip

\item{1)} the semimodule $V$ has the b-approximation property;
\smallskip

\item{2)} every b-linear mapping from $V$ to an arbitrary b-complete
semimodule $W$ over $K$ is b-nuclear;
\smallskip

\item{3)} every b-linear mapping from an arbitrary b-complete
semimodule $W$ over $K$ to the semimodule $V$ is b-nuclear.
\end{proposition}

\begin{corollary} An arbitrary b-complete semimodule over
a b-complete semiring $K$ is b-nuclear if and only if this
semimodule has the b-approximation property.
\end{corollary}

Recall that, in the traditional functional analysis, any
nuclear space has the approximation property but the converse
is not true.

Concrete examples of $b$-nuclear spaces and semimodules are
described in Examples 7.15, 7.16 and 7.19 (see above). Important $b$-nuclear spaces and semimodules (e.g., the so-called Lipschitz spaces and
semi-Lipschitz semimodules) are described in \cite{LiSh2007}.
In this paper there is a description of all functional $b$-semimodules
for which the kernel theorem holds (as semi-Lipschitz semimodules);
this result is due to G.~B.~Shpiz.

It is
easy to show that the idempotent spaces $USC(X)$ and
Conc($\R$)  (see Examples 7.17 and 7.18) are not $b$-nuclear
(however, for these spaces the kernel theorem is true). The reason
is that these spaces are not functional $b$-spaces and the
corresponding $\delta$-functionals are not $b$-linear
(and even linear).
\medskip

\subsection{Kernel theorems for functional $b$-semimodules}

Let $V\subset K(X)$ be a $b$-complete functional semimodule
over a $b$-complete semiring $K$. Recall that for $V$ the
{\it kernel theorem} holds if every $b$-linear mapping of
this semimodule to an arbitrary $b$-complete semimodule over
$K$ has an integral representation.

\begin{theorem} Assume that  a b-complete
semiring $K$ and a nonempty set $X$ are given. The kernel theorem
holds for any functional b-semimodule $V\subset K(X)$ if
and only if every b-linear functional on $V$ is integral
and the semimodule $V$ is b-nuclear, i.e., has the
b-approximation property.
\end{theorem}

\begin{corollary} If for a functional b-semimodule
the kernel theorem holds, then this semimodule is b-nuclear.
\end{corollary}

Note that the possibility to obtain an integral representation of
a functional means that one can decompose it into a sum
of functionals of the form $\delta _x$.

\begin{corollary} Assume that  a b-complete semiring $K$
and a nonempty set $X$ are given. The kernel theorem holds for a
functional b-semimodule
$V\subset K(X)$ if and only if the identity operator
id: $V\to V$ is integral.
\end{corollary}
\medskip

\subsection{Integral representations of operators in
 abstract idempotent semimodules}

In this subsection, we examine the following problem: when
a $b$-complete idempotent semimodule $V$ over a $b$-complete
semiring is isomorphic to a functional $b$-semimodule
$W$ such that the kernel theorem holds for $W$.

Assume that $V$ is a $b$-complete idempotent
semimodule over a $b$-complete semiring $K$ and $\phi$ is a
$b$-linear functional defined on $V$. We call this
functional a $\delta$-{\it functional} if there exists an
element $v\in V$ such that
$$
\phi(w)\odot v < w
$$
for every element $w\in V$. It is easy to see that every
functional of the form $\delta_x$ is a $\delta$-functional
in this sense (but the converse is not true in general).

Denote by $\Delta(V)$ the set of all $\delta$-functionals on
$V$. Denote by $i_\Delta$ the natural mapping
$V\to K(\Delta(V))$ defined by the formula
$$
(i_\Delta (v) )(\phi)=\phi(v)
$$
for all $\phi \in \Delta(V)$. We say that an element $v\in V$ is
{\it pointlike} if there exists a $b$-linear functional $\phi$
such that $\phi(w)\odot v < w$ for all $w\in V$. The set of all
pointlike elements of $V$ will be denoted by $P(V)$. Recall that
by $\phi\odot v$ we denote the one-dimensional operator $w\mapsto
\phi(w)\odot v$.

The following assertion is an
obvious consequence of our definitions (including the
definition of the standard order) and the idempotency
of our addition.

\begin{remark} If a one-dimensional operator $\phi\odot v$
appears in the decomposition of the identity operator on $V$
into a sum of one-dimensional operators, then $\phi\in\Delta(V)$
and $v\in P(V)$.
\end{remark}
\medskip
Denote by $id$ and $Id$ the identity operators on $V$ and
$i_\Delta (V)$, respectively.

\begin{proposition} If the operator id is
b-nuclear, then $i_\Delta$ is an embedding and the
operator Id is integral.

If the operator  $i_\Delta$ is an embedding and
the operator Id is integral, then the operator id is
$b$-nuclear.
\end{proposition}

\begin{theorem} A b-complete idempotent semimodule $V$ over
a b-complete idempotent semiring $K$ is isomorphic to a
functional b-semimodule for which the kernel theorem holds
if and only if the identity mapping on $V$ is a
b-nuclear operator, i.e., $V$ is a b-nuclear semimodule.
\end{theorem}

The following proposition shows that, in a certain sense, the
embedding $i_\Delta$ is a universal representation of a
$b$-nuclear semimodule in the form of a functional
$b$-semimodule for which the kernel theorem holds.

\begin{proposition} Let $K$ be a b-complete
idempotent semiring, $X$ be a nonempty set, and $V\subset
K(X)$ be a functional $b$-semimodule on $X$ for which the
kernel theorem holds. Then there exists a natural mapping
$i:X\to \Delta(V)$ such that the corresponding mapping
$i_*: K(\Delta(V))\to K(X)$ is an isomorphism of
$i_\Delta(V)$ onto~$V$.
\end{proposition}
\medskip

\section{The dequantization transform, convex geometry and the Newton polytopes}

Let $X$ be a topological space. For functions $f(x)$ defined on
$X$ we shall say that a certain property is valid {\it almost
everywhere} (a.e.) if it is valid for all elements $x$ of an open
dense subset of $X$. Suppose $X$ is $\C^n$ or $\R^n$; denote by
$\R^n_+$ the set $x=\{\,(x_1, \dots, x_n)\in X \mid x_i\geq 0$ for
$i = 1, 2, \dots, n$.
 For $x= (x_1, \dots, x_n) \in X$ we set
${\mbox{exp}}(x) = ({\mbox{exp}}(x_1), \dots, {\mbox{exp}}(x_n))$;
so if $x\in\R^n$, then ${\mbox{exp}}(x)\in \R^n_+$.

Denote by $\cF(\C^n)$ the set of all functions defined and
continuous on an open dense subset $U\subset \C^n$ such that
$U\supset \R^n_+$. It is clear that $\cF(\C^n)$ is a ring (and an
algebra over $\C$) with respect to the usual addition and
multiplications of functions.

For $f\in \cF(\C^n)$ let us define the function $\hat f_h$ by the
following formula:
$$
\hat f_h(x) = h \log|f({\mbox{exp}}(x/h))|,
\eqno(14)
$$
where $h$ is a (small) real positive parameter and $x\in\R^n$. Set
$$
\label{e:hatfx} \hat f(x) = \lim_{h\to +0} \hat f_h (x),
\eqno(15)
$$
if the right-hand side of (15) exists almost everywhere.

We shall say that the function $\hat f(x)$ is a {\it
dequantization} of the function $f(x)$ and the map $f(x)\mapsto
\hat f(x)$ is a {\it dequantization transform}. By construction,
$\hat f_h(x)$ and $\hat f(x)$ can be treated as functions taking
their values in $\R_{\max}$. Note that in fact $\hat f_h(x)$ and
$\hat f(x)$
 depend on the restriction of $f$ to $\R_+^n $
only; so in fact the dequantization transform is constructed for
functions defined on $\R^n_+$ only. It is clear that the
dequantization transform is generated by the Maslov dequantization
and the map $x\mapsto |x|$.

Of course, similar definitions can be given for functions defined
on $\R^n$ and $\R_+^n$. If $s=1/h$, then we have the following
version of (14) and (15):
$$
\hat f(x) = \lim_{s\to \infty} (1/s) \log|f(e^{sx})|.
\eqno(15')
$$

Denote by $\partial \hat f$ the subdifferential of the function
$\hat f$ at the origin.

If $f$ is a polynomial  we have
$$
\partial \hat f = \{\, v\in \R^n\mid (v, x) \le \hat f(x)\
\forall x\in \R^n\,\}.
$$
It is well known that all the convex compact subsets in $\R^n$
form an idempotent semiring $\mathcal{S}$ with respect to the
Minkowski operations: for $\alpha, \beta \in \mathcal{S}$ the sum
$\alpha\oplus \beta$ is the convex hull of the union $\alpha\cup
\beta$; the product $\alpha\odot \beta$ is defined in the
following way: $\alpha\odot \beta = \{\, x\mid x = a+b$, where
$a\in \alpha, b\in \beta$, see Fig.3. In fact $\mathcal{S}$ is an
idempotent linear space over $\R_{\max}$.

Of course, the Newton polytopes of polynomials in $n$ variables
form a subsemiring
$\mathcal{N}$ in $\mathcal{S}$. If $f$, $g$ are polynomials, then
$\partial(\widehat{fg}) = \partial\hat f\odot\partial\widehat g$;
moreover, if $f$ and $g$ are ``in general position'', then
$\partial(\widehat{f+g}) = \partial\hat f\oplus\partial\widehat
g$. For the semiring of all polynomials with nonnegative
coefficients the dequantization transform is a homomorphism of
this ``traditional'' semiring to the idempotent semiring
$\mathcal{N}$.

\begin{figure}
\centering \epsfig{file=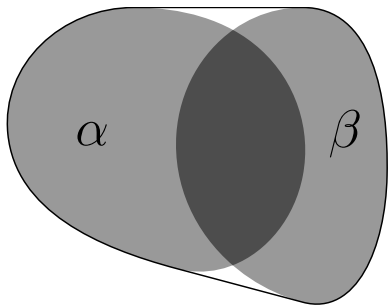,width=0.4\linewidth}
\caption{Algebra of convex subsets.}
\end{figure}

\begin{theorem}
If $f$ is a polynomial, then the subdifferential $\partial\hat f$
of $\hat f$ at the origin coincides with the Newton polytope of
$f$.  For the semiring of polynomials with nonnegative
coefficients, the transform $f\mapsto\partial\hat f$ is a
homomorphism of this semiring to the semiring of convex polytopes
with respect to the Minkowski operations (see above).
\end{theorem}

Using the dequantization transform it is possible to generalize
this result to a wide class of functions and convex
sets, see below and ~\cite{LiSh2005}.
\medskip

\subsection{Dequantization transform: algebraic properties}

Denote by $V$ the set $\rset^n$ treated as a linear Euclidean
space (with the scalar product $(x, y) = x_1y_1+ x_2y_2 +\dots + x_ny_n$)
and set $V_+ = \rset_+^n$.
We shall say that a function $f\in \maF(\cset^n)$ is {\it dequantizable}
whenever its
dequantization $\hat f(x)$ exists (and is defined on an open dense subset
of $V$). By $\maD (\cset^n)$ denote the set of all dequantizable functions and by
$\widehat{\maD}(V)$ denote the set $\{\,\hat f \mid f\in \maD(\cset^n)\,\}$. Recall that
functions from $\maD(\cset^n)$ (and $\widehat{\maD}(V)$) are defined almost everywhere and
$f=g$ means that $f(x) = g(x)$ a.e., i.e., for $x$ ranging over an open dense subset
of $\cset^n$ (resp., of $V$). Denote by $\maD_+(\cset^n)$ the set of all
functions $f\in \maD(\cset^n)$
such that $f(x_1, \dots, x_n)\geq 0$ if $x_i\geq 0$ for $i= 1,\dots, n$; so
 $f\in \maD_+(\cset^n)$ if the restriction of $f$ to $V_+ = \rset_+^n$ is a
 nonnegative function. By $\widehat{\maD}_+(V)$ denote the image of $\maD_+(\cset^n)$
 under the dequantization transform. We shall say that functions
$f, g\in \maD(\cset^n)$ are in {\it general position} whenever
$\hat f (x) \neq \widehat g(x)$ for $x$ running an open dense
subset of $V$.

\begin{theorem}
For functions $f, g \in \maD(\cset^n)$ and any nonzero constant $c$, the
following equations are valid:

\begin{enumerate}
\item[1)] $\widehat{fg} = \hat f + \widehat g$;
\item[2)] $|\hat f| = \hat f$; $\widehat{cf} = f$; $\widehat c =0$;
\item[3)] $(\widehat{f+g})(x) = \max\{\hat f(x), \widehat g(x)\}$ a.e.\ if $f$ and $g$
are nonnegative on $V_+$ (i.e., $f, g \in \maD_+(\cset^n)$) or $f$ and $g$
 are in general position.
\end{enumerate}
Left-hand sides of these equations are well-defined automatically.
\end{theorem}

\begin{corollary}
The set $\maD_+(\cset^n)$ has a natural structure of a semiring with respect to
the
 usual addition and multiplication of functions taking their values in $\cset$.
 The set $\widehat{\maD}_+(V)$ has a natural
 structure of an idempotent semiring with respect to the operations
$(f\oplus g)(x) = \max \{ f(x), g(x)\}$, $(f\odot g)(x) = f(x) + g(x)$;
elements
 of $\widehat{\maD}_+(V)$ can be naturally treated as functions taking their values
 in $\rset_{\max}$. The dequantization transform generates a homomorphism from
 $\maD_+(\cset^n)$ to $\widehat{\maD}_+(V)$.
\end{corollary}
\medskip

\subsection{Generalized polynomials and simple functions}

For any nonzero number $a\in\cset$ and any vector
 $d = (d_1, \dots, d_n)\in V = \rset^n$
we set $m_{a,d}(x) = a \prod_{i=1}^n x_i^{d_i}$; functions of this kind we
 shall
call {\it generalized monomials}. Generalized monomials are defined a.e.\ on
$\cset^n$ and on $V_+$, but not on $V$ unless the numbers $d_i$ take
integer or suitable rational values. We shall say that a function $f$ is a {\it
generalized polynomial} whenever it is a finite sum of linearly
 independent generalized monomials. For instance, Laurent polynomials
and Puiseax polynomials are examples of generalized polynomials.

As usual, for $x, y\in V$ we set $(x,y) = x_1y_1 + \dots + x_ny_n$. The
following proposition is a result of a trivial calculation.

\begin{proposition}
For any nonzero number $a\in V = \cset$ and any vector $d\in V = \rset^n$
we have $(\widehat{m_{a,d}})_h(x) = (d, x) + h\log|a|$.
\end{proposition}

\begin{corollary}
If $f$ is a generalized monomial, then $\hat f$ is a linear function.
\end{corollary}

Recall that a real function $p$ defined on $V = \rset^n$ is {\it sublinear}
if $p = \sup_{\alpha}
p_{\alpha}$, where $\{p_{\alpha}\}$ is a collection of linear functions.
Sublinear functions defined everywhere on $V=\rset^n$ are convex; thus these
 functions are continuous, see \cite{MaTi2003}.
 We discuss sublinear functions of this kind only. Suppose $p$ is a continuous
 function defined on $V$, then $p$ is sublinear whenever

1) $p(x+ y) \leq p(x) + p(y)$ for all $x, y \in V$;

2) $p(cx) = cp(x)$ for all $x\in V$, $c\in \rset_+$.

So if $p_1$, $p_2$ are sublinear functions, then $p_1 +p_2$ is a sublinear
 function.

We shall say
that a function $f \in \maF(\cset^n)$ is {\it simple}, if its dequantization $\hat f$
exists and a.e.\ coincides with a sublinear function; by misuse of language, we
shall denote this (uniquely defined everywhere on $V$) sublinear
 function by the same symbol $\hat f$.

 Recall that simple functions $f$ and $g$ are {\it in general position} if
$\hat f(x) \neq \widehat g(x)$ for all $x$ belonging to an open dense subset of $V$.
In particular, generalized monomials are in
general position whenever they are linearly independent.

Denote by $\mathit{Sim}(\cset^n)$ the set of all simple functions defined on $V$ and
denote by $\mathit{Sim}_+(\cset^n)$ the set $\mathit{Sim}(\cset^n) \cap \maD_+(\cset^n)$. By
$\mathit{Sbl}(V)$ denote the
 set of all (continuous) sublinear functions defined on $V = \rset^n$ and by
 $\mathit{Sbl}_+(V)$ denote the image $\widehat{\mathit{Sim}_+}(\cset^n)$ of $\mathit{Sim}_+(\cset^n)$ under the
 dequantization transform.

The following statements can be easily deduced from Theorem 8.2 and definitions.

\begin{corollary}
The set $\mathit{Sim}_+(\cset^n)$ is a subsemiring of $\maD_+(\cset^n)$ and $\mathit{Sbl}_+(V)$
is an idempotent subsemiring of $\widehat{\maD_+}(V)$. The
dequantization transform generates an epimorphism of
$\mathit{Sim}_+(\cset^n)$ onto $\mathit{Sbl}_+(V)$. The set $\mathit{Sbl}(V)$ is an idempotent
semiring with respect to the operations
$(f\oplus g)(x) = \max \{ f(x), g(x)\}$,
$(f\odot g)(x) = f(x) + g(x)$.
\end{corollary}

\begin{corollary}
Polynomials and generalized polynomials are simple functions.
\end{corollary}

We shall say that functions $f, g\in\maD(V)$ are {\it asymptotically equivalent}
whenever $\hat f = \widehat g$; any simple function $f$ is an {\it asymptotic
monomial} whenever
$\hat f$ is a linear function. A simple function $f$ will be called an {\it
asymptotic polynomial} whenever $\hat f$ is a sum of a finite collection of
nonequivalent asymptotic monomials.

\begin{corollary}
Every asymptotic polynomial is a simple function.
\end{corollary}

\begin{example} Generalized polynomials, logarithmic functions of
(generalized) polynomials, and products of
polynomials and logarithmic functions are asymptotic polynomials. This follows
from our definitions and formula~(15).
\end{example}
\medskip

\subsection{Subdifferentials of sublinear functions}

We shall use some elementary results from convex analysis. These results can be
found, e.g., in \cite{MaTi2003}, ch. 1, \S 1.

For any function $p\in \mathit{Sbl}(V)$ we set
$$
\partial p = \{\, v\in V\mid (v, x) \le p(x)\ \forall x\in V\,\}.
$$

It is well known from convex analysis that for any sublinear function $p$ the
set $\partial p$ is exactly the {\it subdifferential} of $p$ at the origin.
 The following propositions are also known in convex
analysis.

\begin{proposition}
Suppose $p_1,p_2\in \mathit{Sbl}(V)$, then
\begin{enumerate}
\item[1)] $\partial (p_1+p_2) = \partial p_1\odot\partial p_2 = \{\, v\in V\mid
v = v_1+v_2, \text{ where $v_1\in \partial p_1, v_2\in \partial p_2$}\,\}$;
\item[2)] $\partial (\max\{p_1(x), p_2(x)\}) = \partial p_1\oplus\partial p_2$.
\end{enumerate}
\end{proposition}

Recall that $\partial p_1\oplus \partial p_2$ is a convex hull of the set
$\partial p_1\cup \partial p_2$.

\begin{proposition}
Suppose $p\in \mathit{Sbl}(V)$.  Then $\partial p$ is a nonempty convex compact
subset of $V$.
\end{proposition}

\begin{corollary}
The map $p\mapsto \partial p$ is a homomorphism of the idempotent semiring
 $\mathit{Sbl}(V)$ (see Corollary~8.3) to the idempotent semiring $\mathcal{S}$ of all convex
 compact subsets of $V$ (see Subsection~8.1 above).
\end{corollary}
\medskip

\subsection{Newton sets for simple functions}

For any simple function $f\in \mathit{Sim}(\cset^n)$ let us denote by $N(f)$ the set
 $\partial(\hat f)$. We shall call $N(f)$ the {\it Newton set} of the
 function $f$.

\begin{proposition}
For any simple function $f$, its Newton set $N(f)$ is a nonempty convex
 compact subset of $V$.
\end{proposition}

This proposition follows from Proposition 8.11 and definitions.

\begin{theorem}
Suppose that $f$ and $g$ are simple functions. Then
\begin{enumerate}
\item[1)] $N(fg) = N(f)\odot N(g) = \{\, v\in V\mid v = v_1 +v_2$
 with
 $v_1 \in N(f), v_2 \in N(g)$;
\item[2)] $N(f+g) = N(f)\oplus N(g)$, if $f_1$ and $f_2$ are in general
 position or $f_1, f_2 \in \mathit{Sim}_+(\cset^n)$ {\rm (}recall that $N(f)\oplus N(g)$
is the convex hull of $N(f)\cup N(g)${\rm )}.
\end{enumerate}
\end{theorem}

This theorem follows from Theorem~8.2, Proposition~8.10 and definitions.

\begin{corollary}
The map $f\mapsto N(f)$ generates a homomorphism from $\mathit{Sim}_+(\cset^n)$ to
$\mathcal{S}$.
\end{corollary}

\begin{proposition}
Let $f = m_{a,d}(x) = a \prod^n_{i=1} x_i^{d_i}$ be a monomial; here
$d = (d_1, \dots, d_n) \in V= \rset^n$ and $a$ is a nonzero
complex number. Then $N(f) = \{ d\}$.
\end{proposition}

This follows from Proposition 8.4, Corollary 8.5 and definitions.

\begin{corollary}
Let $f = \sum_{d\in D} m_{a_d,d}$ be a polynomial. Then $N(f)$ is the polytope
$\oplus_{d\in D}\{d\}$, i.e.\ the convex hull of the finite set $D$.
\end{corollary}

This statement follows from Theorem 8.14 and Proposition 8.16. Thus in this case
 $N(f)$ is the well-known classical Newton polytope of the polynomial $f$.

Now the following corollary is obvious.

\begin{corollary}
Let $f$ be a generalized or asymptotic polynomial. Then its Newton set
 $N(f)$ is a convex polytope.
\end{corollary}

\begin{example} Consider the one dimensional case, i.e., $V = \rset$ and
suppose
 $f_1 = a_nx^n + a_{n-1}x^{n-1} + \dots + a_0$ and $f_2 = b_mx^m + b_{m-1}
 x^{m-1} + \dots + b_0$, where $a_n\neq 0$, $b_m\neq 0$, $a_0 \neq 0$,
 $b_0 \neq 0$. Then $N(f_1)$ is the segment $[0, n]$ and $N(f_2)$ is the
 segment $[0, m]$. So the map $f\mapsto N(f)$ corresponds to the map
 $f\mapsto \deg (f)$, where $\deg(f)$ is a degree of the polynomial $f$. In
 this case Theorem 2 means that $\deg(fg) = \deg f + \deg g$ and
 $\deg (f+g) = \max \{\deg f, \deg g\} = \max \{n, m\}$ if $a_i\geq 0$,
 $b_i\geq 0$ or $f$ and $g$ are in general position.
\end{example}
\medskip

\section{Dequantization of set functions and measures on metric spaces}
\medskip

The following results are presented in~\cite{LiSh}.

\begin{example} Let $M$ be a metric space, $S$ its arbitrary
subset with a compact closure. It is well-known that a Euclidean
$d$-dimensional ball $B_{\rho}$ of radius $\rho$ has volume
$$
\ovol\nolimits_d(B_{\rho})=\frac{\Gamma(1/2)^d}{\Gamma(1+d/2)}\rho^d,
$$
where $d$ is a natural parameter. By means of this formula it is
possible to define a volume of $B_{\rho}$ for any {\it real} $d$.
Cover $S$ by a finite number of balls of radii $\rho_m$. Set
$$
v_d(S):=\lim_{\rho\to 0} \inf_{\rho_m<\rho} \sum_m
\ovol\nolimits_d(B_{\rho_m}).
$$
Then there exists a number $D$ such that $v_d(S)=0$ for $d>D$ and
$v_d(S)=\infty$ for $d<D$. This number $D$ is called the {\it
Hausdorff-Besicovich dimension} (or {\it HB-dimension}) of $S$,
see, e.g.,~\cite{Ma2005}. Note that a set of non-integral
HB-dimension is called a fractal in the sense of B.~Mandelbrot.
\end{example}

\begin{theorem}
Denote by $\cN_{\rho}(S)$ the minimal number of balls of radius
$\rho$ covering $S$. Then
$$
D(S)=\mathop{\underline{\lim}}\limits_{\rho\to +0} \log_{\rho}
(\cN_{\rho}(S)^{-1}),
$$
where $D(S)$ is the HB-dimension of $S$. Set $\rho=e^{-s}$, then
$$
D(S)=\mathop{\underline{\lim}}\limits_{s\to +\infty} (1/s) \cdot
\log \cN_{exp(-s)}(S).
$$
So the HB-dimension $D(S)$ can be treated as a result of a
dequantization of the set function $\cN_{\rho}(S)$.
\end{theorem}

\begin{example} Let $\mu$ be a set function on $M$ (e.g., a
probability measure) and suppose that $\mu(B_{\rho})<\infty$ for
every ball $B_{\rho}$. Let $B_{x,\rho}$ be a ball of radius $\rho$
having the point $x\in M$ as its center. Then define
$\mu_x(\rho):=\mu(B_{x,\rho})$ and let $\rho=e^{-s}$ and
$$
D_{x,\mu}:=\mathop{\underline{\lim}}\limits_{s\to +\infty}
-(1/s)\cdot\log (|\mu_x(e^{-s})|).
$$
This number could be treated as a dimension of $M$ at the point
$x$ with respect to the set function $\mu$. So this dimension is a
result of a dequantization of the function $\mu_x(\rho)$, where
$x$ is fixed. There are many dequantization procedures of this
type in different mathematical areas. In particular, V.P.~Maslov's
negative dimension (see~\cite{Mas2007}) can be treated similarly.
\end{example}
\medskip

\section{Dequantization of geometry}

An idempotent version of real algebraic geometry was discovered in
the report of O.~Viro for the Barcelona Congress~\cite{Vir2000}.
Starting from the idempotent correspondence principle O.~Viro
constructed a piecewise-linear geometry of polyhedra of a special
kind in finite dimensional Euclidean spaces as a result of the
Maslov dequantization of real algebraic geometry. He indicated
important applications in real algebraic geometry (e.g.,
 in the framework of Hilbert's 16th problem for
constructing real algebraic varieties with prescribed properties
and parameters) and relations to complex algebraic geometry and
amoebas in the sense of I.~M.~Gelfand, M.~M.~Kapranov, and
A.~V.~Zelevinsky, see~\cite{GeKaZe94,Vir2002}. Then complex
algebraic geometry was dequantized by G.~Mikhalkin and the result
turned out to be the same; this new `idempotent' (or asymptotic)
geometry is now often called the {\it tropical algebraic
geometry}, see,
e.g.,~\cite{Iten2007,LiMa2005,LiSer2007,LiSer2009,Mi2005,Mi2006}.

There is a natural relation between the Maslov dequantization and
amoebas.

Suppose $({\cset}^*)^n$ is a complex torus, where ${\cset}^* =
{\cset}\backslash \{0\}$ is the group of nonzero complex numbers
under multiplication.  For
 $z = (z_1, \dots, z_n)\in
(\cset^*)^n$ and a positive real number $h$ denote by $\Log_h(z) =
h\log(|z|)$ the element
\[(h\log |z_1|, h\log |z_2|, \dots,
h\log|z_n|) \in \rset^n.\] Suppose $V\subset (\cset^*)^n$ is a
complex algebraic variety; denote by $\maA_h(V)$ the set
$\Log_h(V)$. If $h=1$, then the set $\maA(V) = \maA_1(V)$ is
called the {\it amoeba} of $V$; the amoeba $\maA(V)$ is a closed
subset of $\rset^n$ with a non-empty complement. Note that this
construction depends on our coordinate system.

For the sake of simplicity suppose $V$ is a hypersurface
in~$(\cset^*)^n$ defined by a polynomial~$f$; then there is a
deformation $h\mapsto f_h$ of this polynomial generated by the
Maslov dequantization and $f_h = f$ for $h = 1$. Let $V_h\subset
({\cset}^*)^n$ be the zero set of $f_h$ and set $\maA_h (V_h) =
{\Log}_h (V_h)$. Then
 there exists a tropical variety
$\mathit{Tro}(V)$ such that the subsets $\maA_h(V_h)\subset
\rset^n$ tend to $\mathit{Tro}(V)$ in the Hausdorff metric as
$h\to 0$. The tropical variety $\mathit{Tro}(V)$ is a result of a
deformation of the amoeba $\maA(V)$ and the Maslov dequantization
of the variety $V$. The set $\mathit{Tro}(V)$ is called the {\it
skeleton} of $\maA(V)$.

\begin{figure}
\noindent\epsfig{file=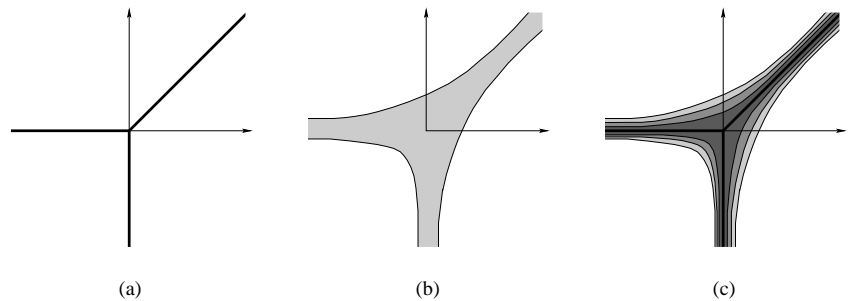,width=0.9\linewidth}
\caption{Tropical line and deformations of an amoeba.}
\end{figure}

\begin{example}  For the line $V = \{\, (x, y)\in ({\cset}^*)^2
\mid x + y + 1 = 0\,\}$ the piecewise-linear graph
$\mathit{Tro}(V)$ is a tropical line, see Fig.4(a). The amoeba
$\maA(V)$ is represented in Fig.4(b), while Fig.4(c) demonstrates
the corresponding deformation of the amoeba.
\end{example}

\medskip

{\bf Acknowledgments}.  The author is sincerely grateful to V.~N.~Kolokoltsov,
V.~P.~Maslov, G.~B.~Shpiz, S.~N.~Sergeev, A.~N.~Sobolevski{{\u\i}}, and A.~V.~Tchourkin
for valuable suggestions, help and support.
\medskip

\end{document}